# Micro-coil detection of Nuclear Magnetic Resonance for nanofluidic samples

A. Shibahara[1], A. Casey[1], C.P. Lusher[1], J. Saunders[1], C. Aßmann[2], Th. Schurig[2] and D. Drung[2]

[1]Department of Physics, Royal Holloway University of London, Egham, Surrey, TW20 0EX, United Kingdom

[2]Physikalisch-Technische Bundesanstalt, Abbestrasse 2-12, D-10587 Berlin, Germany

We have developed a novel dc SQUID system with a micro-coil input circuit to act as a local probe of quantum matter and nanosystems. The planar niobium micro-coil pickup loop is located remotely from the SQUID, coupled through a superconducting twisted pair, enabling the sample to be at microkelvin temperatures. A high degree of coupling between the coil and the region of interest of similar dimensions (up to ~ 100 microns) can be achieved. We report nuclear magnetic resonance (NMR) measurements to characterise the sensitivity of these coils to $^3$He in the gas phase at 4.2 K in a 30 mT magnetic field.

The study of nano-samples near their quantum mechanical ground-state operating at ultralow temperatures poses experimental challenges to achieve sample thermalization and requires measurement techniques with extremely low levels of heating. An inductively coupled measurement, avoiding the direct contact of wires and measurement leads to the sample, can reduce the parasitic heat input by orders of magnitude. The NMR response of mesoscopic scale objects provides an important non-invasive experimental probe of such systems. The intrinsic sensitivity of superconducting quantum interference devices[1], SQUIDs, to magnetic flux makes these devices ideally suited for these studies.

In conventional NMR it has been demonstrated that improved sensitivity can be achieved by reducing the size of the pickup loop to match the dimensions of the region of interest[2]. In commercial NMR spectroscopy, microlitre samples are routinely studied in the analysis of volume limited samples, using solenoidal-microcoil based probes. Advances in these techniques[3,4,5,6] have enabled scientists to characterise nanolitre volume samples.

Miniature magnetometers[7] and susceptometers[8,9], in which particles are placed directly over the SQUID loop or integrated pickup coil have been used in the study of the properties of semiconductors and superconductors. In a similar manner NMR has been carried out on a 50 µm platinum particle.[10] Sub-micron and nanoscale SQUIDs[11] and SQUIDs with an integrated nanoloop[12] have been fabricated for the study of the magnetic properties of small isolated samples, with the ultimate aim of single spin detection.

Improved spatial resolution has also been achieved through employing magnetic resonance imaging techniques[13] with microcoils and the application of large field gradients. A high resolution magnetic resonance imaging microscope is under development for the study of $^3$He at ultra-low temperatures[14], which uses a static magnetic field of 7.2T and tri-axial magnetic field gradients of 2 T/m. A combination of miniature SQUID susceptometers with piezoelectric nanopositioning devices has been used for the study of weak magnetic signals from mesoscopic objects[15].

In this Letter we report the detection of NMR signals using microcoils coupled to a remote SQUID detector. This confers significant additional flexibility on the possible sample environment (e.g. in terms of applied field and temperature) and avoids issues associated with power dissipation in the SQUID. The particular application of interest to us is the study of nanofluidic samples of superfluid $^3$He. It has been shown that controlled confinement of nanofluidic[16] samples provides a new laboratory for the study of topological superfluids and their surface and edge bound excitations. Nano-scale samples of quantum fluids provide clean model systems for addressing problems of fundamental significance and technological relevance. NMR provides a tool to identify the order parameter[17,18] of the superfluid within the confined geometry, however identifying point defects, domain walls[19] or the phase at well-defined positions within the cavity provides an experimental challenge. The phases of interest only occur at sub millikelvin temperatures; the volume of the region of interest may be as low as picolitres; applying large fields would significantly perturb the superfluid order parameter under investigation and the helium needs to be confined within walls at least 100 microns thick. This letter reports the results of tests on planar niobium microcoils coupled to a low noise dc SQUID designed to overcome these difficulties. Earlier studies of $^3$He with planar microcoils used a less sensitive GaAs MESFET preamplifier[20]. Signals from 1 bar of $^3$He gas cooled to 4.2 K were obtained to test the calculated signal-to-noise ratio (SNR) and determine the suitability of the microcoils for pulsed NMR measurements. These measurements enable a prediction of the possible SNR for a picolitre volume of superfluid $^3$He.

Two designs of microcoil were tested, a simple single layer design and a more complicated design that combined optical (ultra-violet, UV) and electron-beam lithography predicted to improve the SNR. Both microcoil designs, shown schematically in figure 1a, have an outer square loop size of 500 µm x 500 µm ending in two bonding pads 3.5 mm away. The single layer design consists of a 10 turn pickup coil with a 2.5 µm linewidth, 5 µm pitch and internal loop size 410 µm x 410 µm. The coil is made using only UV lithography and is relatively simple to pattern. The constraint that lines may not cross over results in the pickup coil having an inductance of 105 nH with significant stray inductance in the bonding pads, such that the total inductance at the bonding pads is approximately 400 nH. The calculated field for unit current in the pickup coil, which determines the sensitivity of the coil by the principle of reciprocity[2], is shown in figure 1b. The dimensions of the coil were chosen such that the introduction of a 100 micron thick wall; required for the nanofluidic samples, would not dramatically reduce the sensitivity. Calculations are presented for the axial field for a square pickup coil, using the Biot-Savart law, with the approximation that the turns are perfect squares. The profile in figure 1b determines the depth of sample that will be assayed by the coil.

With the trilayer design where both UV and e-beam lithography have been used, crossovers of the patterned lines become possible. Figure 1c shows a close up of the crossover region showing how the two ends of the coil are joined to the lines leading to the output pads. With e-beam lithography, a linewidth of between 600 and 800 nm is obtained, with a 2 µm pitch. The pickup coil has 18 turns, an internal loop size of 432 µm x 432 µm, with an increased inductance of 394 nH. With minimal stray inductance, the total inductance at the bonding pads remains at ~400 nH.

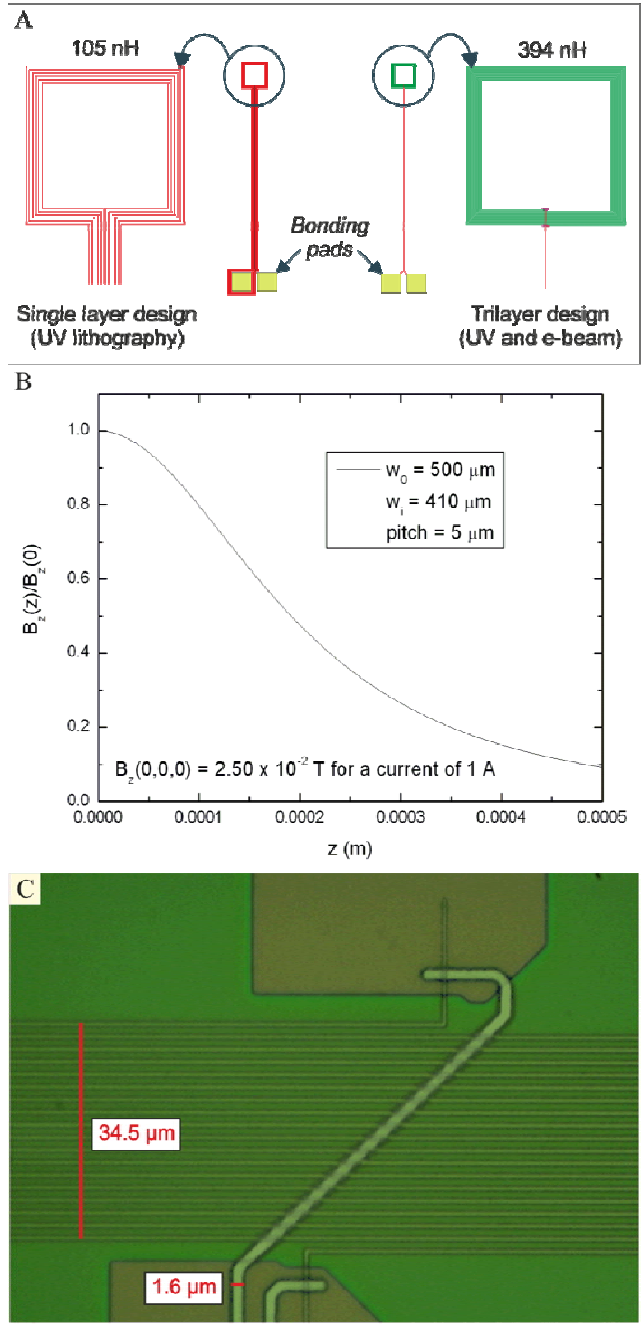

FIG. 1. a) Schematic of the two microcoil designs, b) Calculated axial field profile of the single layer coil as a function of distance, z, along the axis of the microcoil, c) Close up of the crossover section of the trilayer coil. The thicker lines join the two ends of the coil, which is patterned afterwards using e-beam lithography (color online).

We couple the microcoils to a SQUID via a tuned input circuit configuration[21] shown in figure 2, which provides a good match between the receiver coil and the SQUID input coil. This configuration does not require the cold step-up transformer used in earlier work[20]. The optimization condition[21] on resonance at frequency $\omega_0$ is, $\omega_0 M_i^2 / R_i L_s = 1$, where $M_i$ is the mutual inductance between the input coil and the SQUID, $L_s$ is the inductance of the SQUID and $R_i$ is the resistance in the input circuit. This optimization condition can be written in terms of the quality factor of the input circuit, $Q = (\omega_0 L_T / R_i)$, and the coupling constant $\alpha_i^2$ to give, $\alpha_i^2 Q L_i / L_T = 1$, such that the optimum total input inductance is, $L_T = \alpha_i^2 Q L_i$. This is made up of the sum of the pickup coil inductance, the SQUID input coil inductance and any additional stray inductance. PTB Berlin fabricates a series of sensitive two-stage SQUIDs with a wide range of input inductances[22]. For this work a practical target value of the quality factor for NMR experiments is $Q \sim 30$. In this case the low inductance X116 SQUIDs most closely match the optimum condition, with $L_s = 85$ pH. These SQUIDs have a 4 turn input coil with $L_i = 29$ nH including the integrated $Q$-spoiler, with a coupling constant $\alpha_i^2 = 0.36$.

To create the tuned spectrometer the microcoil is connected via a superconducting twisted pair, fabricated from niobium-titanium wire, to a remote SQUID separated by approximately 20 cm, in series with a low loss 47.5 nF copper and Teflon capacitor[23]. The connection to the microcoil was made through aluminium wire bonds to copper solder pads, which conveniently provided the resistive element (of order 0.1 Ω) needed for the required $Q$ at the temperature of the test measurements (4.2 K). From the measured 4.2 K noise peaks, shown in figure 3, and the known capacitance, the total inductance of the microcoil input circuits were $L_T = 617$ nH and 614 nH, for the single and tri-layer coils respectively. This is sufficiently close to the optimum value not to significantly degrade the SNR of the spectrometer. The two coils were designed with the same total inductance for a direct comparison of performance.

The microcoils were each sealed inside a sample chamber made from Stycast 1266 into which $^3$He gas was added as a test sample. The $^3$He gas was therefore in direct contact with one side of the microcoil, which as seen in figure 1b, is sensitive to a region of sample of depth approximately equal to the coil diameter. The sample cell was placed within a saddle transmitter coil orthogonal to the microcoil, inside a solenoid NMR magnet. Thus the static NMR field $B_0$ was orthogonal to both the transmitter coil and the microcoil.

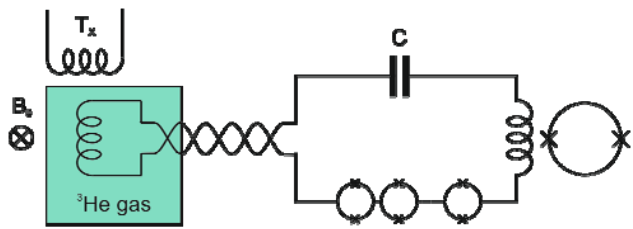

FIG. 2. Schematic of tuned NMR setup. The top surface of the microcoil is exposed to the $^3$He gas sample. All elements are at 4.2 K. The arrangement is housed in a superconducting niobium enclosure to shield from extraneous noise (color online).

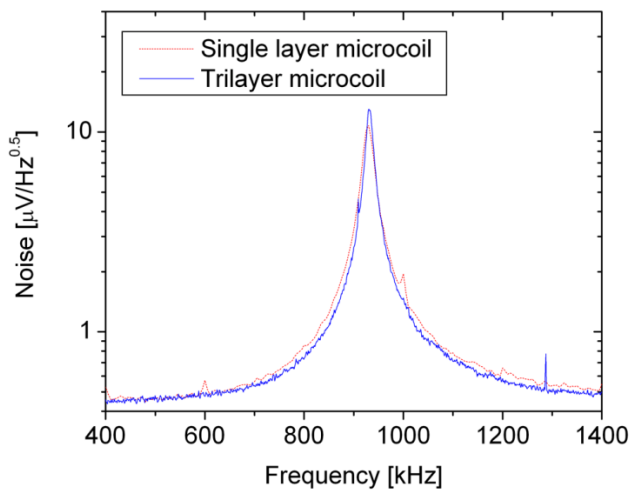

FIG. 3. Noise peaks for both designs of microcoil. Showing that total inductance, $L_T$, is 617 nH and 614 nH for the single layer microcoil and trilayer microcoil respectively (color online).

NMR measurements were carried out on a sample of $^3$He gas at a pressure of 1 bar and 4.2 K. $B_0$, was set at 28.7 mT to match the Larmor frequency of $^3$He to the resonance frequency of the tuned input circuit. A direct comparison of NMR signals from the two microcoils is shown in figure 4. These signals are Fast-Fourier transforms, FFTs, of the average of 500 free induction decays, captured for 15 ms in the time-domain using a NI-PXI 5922 digitizer[24], following the application of a 90° tipping pulse, and taken with an 80 s repetition time. The repetition time was chosen to be five times greater than the longitudinal relaxation time, $T_1$, which was observed to be 15 s. Fits to the FFTs give a spin-spin relaxation time in the applied field, $T_2^*$ of order 2-3 ms. Within the test cell a magneto-acoustic response to the transmitter pulse is visible, the signals displayed in figure 4 and figure 5 have had this response (taken with the NMR field "detuned") subtracted in the time domain before applying the FFT.

The measured NMR response from the gas sample and the higher SNR of the tri-layer coil are shown in figure 4. Estimating the SNR following a 90° tipping pulse using the expressions given in ref [25], we obtain an SNR of 0.57 in a single shot for the single layer microcoil and 1.0 for the trilayer coil (equivalent to 12.7 and 22.4 after 500 averages), in reasonable agreement with the data in Fig. 4, some noise due to imperfect subtraction of magneto acoustic resonances is apparent. The greater number of turns on the trilayer coil is the dominant factor that accounts for the improved SNR.

We mimic the effect of the wall necessary in the confined nanofluidic superfluid experiment by fixing a 100 micron thick Kapton spacer over the coil with Corning vacuum grease to exclude helium from this region. Integrating the axial field that would be produced by unit current in the coil results in a prediction of a 20% decrease in signal size, for a 100 micron spacer, which is in reasonable agreement with the observed signal in figure 5.

These results enable us to predict with some confidence the achievable SNR from superfluid $^3$He confined in nanofluidic cells. Here we are interested in detecting signals in the 1-10 picolitre range corresponding to 100 µm x 100 µm regions of 100 nm to 1 µm thick cavities. We take into account the higher spin density of the liquid sample, the higher nuclear magnetic susceptibility of the degenerate Fermi liquid and the lower noise contribution from the resistive element in the tuned input circuit, which would be located at low mK temperatures. We assume that the energy resolution of SQUIDs cooled to low mK temperatures[26] to be 2 *h*, an improvement over the ~50 *h* observed at 4.2 K (where *h* is Planck's constant). We find that the SNR should be comparable to that of the present set-up for a 10 picolitre sample.

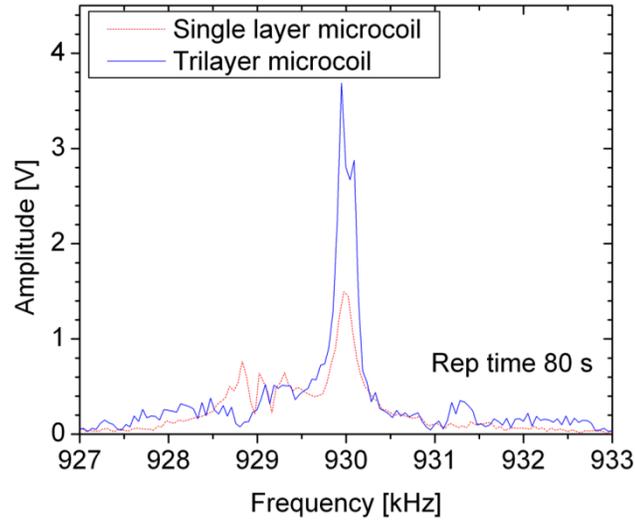

FIG. 4. NMR signals from ³He gas with an 80 s repetition rate, comparing the signal size of a 90° pulse for the two microcoil designs. Each trace consists of 500 averages (color online).

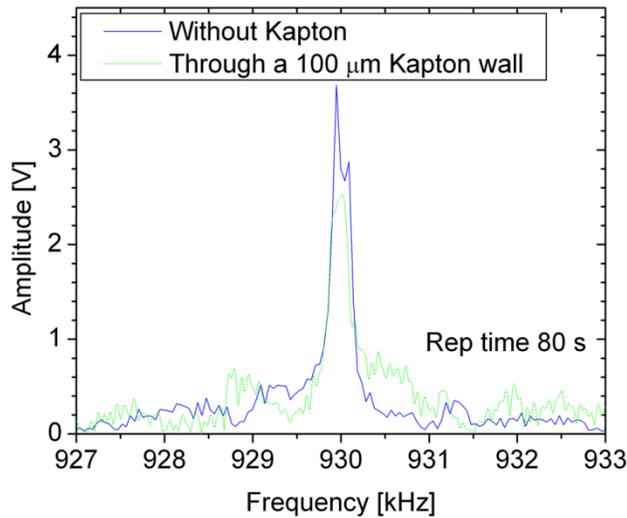

FIG. 5. Reduction in signal size signal size is affected by viewing through a 100 µm thick Kapton wall. Repetition time 80 seconds (color online).

Superconducting Nb planar microcoils coupled to low inductance dc SQUIDs have been tested, configured as a tuned NMR spectrometer. With a trilayer design of microcoil we predict a SNR ~ 4, after 500 averages, for a picolitre volume of superfluid ³He confined in a 100 nm slab, separated from the microcoil by a 100 µm thick cell wall. Microcoils with a larger number of turns fabricated using an improved patterning technology could further enhance the SNR, reducing the required measurement time significantly.

An array of such microcoils coupled to a SQUID in the presence of a magnetic field gradient parallel to the surface can be used to provide spatial resolution over multiple sites in a slab sample. This setup will be used to provide spatial resolution in the search for the NMR response of phase boundaries, surface and edge effects and point defects in the superfluid order parameter of $^3$He. More generally the separation of a dissipative SQUID and the planar microcoil gives the possibility to study the magnetic properties of nanosamples at ultra-low temperatures.

## ACKNOWLEDGMENTS

This work was supported by the MICROKELVIN project. We acknowledge the support of the European Community - Research Infrastructures under the FP7 Capacities Specific Programme, MICROKELVIN project number 228464.